\documentstyle[twoside,fleqn,espcrc2]{article}

\newcommand{\AmS}{{\protect\the\textfont2
  A\kern-.1667em\lower.5ex\hbox{M}\kern-.125emS}}

\title{Gravitational effects on the neutrino oscillation in vacuum}

\author{N. Fornengo $^{\rm a}$\thanks{Presented by N. Fornengo}, C. Giunti 
           \address{Dipartimento di Fisica Teorica,
              Universit\`a di Torino and INFN, Sezione di Torino, 
              via P. Giuria 1, 10125 Torino, Italy},
        C. W. Kim
           \address{Dept. of Physics and Astronomy, 
                The Johns Hopkins University, 
                Baltimore, Maryland 21218, USA 
            and School of Physics, Korea Institute for Advanced Study,
                Seoul 130-012, Korea}
        and J. Song
           \address{Center for Theoretical Physics, 
                Seoul National University, 
                Seoul 151--742, Korea}}
       
\begin{document}

\begin{abstract}
The propagation of neutrinos in a gravitational field is studied by
developing a method of calculating a covariant quantum--mechanical 
phase in a curved space--time. The result is applied 
to neutrino propagation in the Schwarzschild metric.
\end{abstract}

\maketitle

\section{Introduction}

The neutrino oscillation phenomenon
has been discussed in the case of neutrinos propagating
in a flat space--time both in the
plane wave approach for highly relativistic neutrinos \cite{plane}
and in the wave packet formalism \cite{packet}.
In this note we present
a generalization of the plane wave approach to the case of curved 
space--time \cite{ourgrav,Cardall,wudka}. 

In order to generalize the plane--wave approach to the case of a
curved space--time, it is useful to recall the key points which
are used to derive the oscillation probability in the flat
space--time:

\begin{list}{}{\setlength{\rightmargin}{0pt}
    \setlength{\leftmargin}{0pt}\setlength{\itemsep}{2pt}
    \setlength{\topsep}{2pt}}

\item 
$\bullet$ Neutrinos are produced at a space--time point
$A(t_A, \vec x_A)$
as flavor eigenstates 
$|\nu_\alpha \rangle = \sum_k U^*_{\alpha k} |\nu_k\rangle$,
superpositions of the mass eigenstates $|\nu_k\rangle$.
$U$ is the unitary mixing matrix of the neutrino fields.

\item 
$\bullet$ The propagation in the space--time of 
$|\nu_k\rangle$ is described by the quantum--mechanical phase
of a {\em plane wave}. For a neutrino propagating to
$P(t, \vec x)$, the state evolves to
$|\nu_k (t,\vec x; t_A,\vec x_A)\rangle = 
\exp (- i \Phi_k) |\nu_k\rangle $ ,
where
\begin{equation}
\Phi_k = 
E_k \, (t - t_A) - \vec p_k \cdot (\vec x - \vec x_A) \; .
\label{eq:phase}
\end{equation}
The energy $E_k$ and the momentum $\vec p_k$ 
are related by the flat space--time
mass--shell relation $E_k^2 - {\vec p_k}^{\;2} = m_k^2$.

\item 
$\bullet$ In order for the oscillation to occur and to be observed, 
the interference among different mass eigenstates has to
be calculated {\em in a specific space--time point} $B(t_B, \vec x_B)$.
This gives the space--time flavor oscillation probability                     
\begin{eqnarray}
P_{\nu_\alpha \rightarrow \nu_\beta}(t_B,\vec x_B; t_A,\vec x_A) =
\sum_k |U_{\alpha k}|^2 |U_{\beta k}|^2 + & &
\label{eq:pab}
\\
2\; {\mathrm Re} \sum_{k>j} U_{\alpha k}^* U_{\beta k} 
U_{\alpha j} U_{\beta j}^* \exp(-i \Delta\Phi_{kj})\; ,
& &
\nonumber
\end{eqnarray}
where the phase shift $\Delta \Phi_{kj} = \Phi_k - \Phi_j$ is
due to the interference between the $k^{\mathrm th}$ and 
$j^{\mathrm th}$ mass eigenstates.

\item 
$\bullet$ In actual experiments,
the time difference $(t_B - t_A)$ is not measured,
whereas the relative position $|\vec x_B - \vec x_A|$ 
of the source and the detector is known, and an
oscillation probability in space 
$P_{\nu_\alpha \rightarrow \nu_\beta}(\vec x_B; \vec x_A)$ 
can be measured. 
In the plane wave formalism, this can be taken care of 
consistently only for relativistic neutrinos by employing 
the {\em light--ray approximation}, which consists in taking
$(t_B - t_A) = |\vec x_B - \vec x_A|$
in the calculation of the phase $\Phi_k$ of {\em each} mass
eigenstate.
This corresponds to calculate the quantum--mechanical phase
of {\em each} $|\nu_k\rangle$ along the {\em light--ray}
path that links $A$ to $B$. 
In this approximation, Eq.(\ref{eq:phase}) becomes
\begin{equation}
\Phi_k^{\mathrm L} = (E_k - |\vec p_k|) |\vec x_B - \vec x_A|\; .
\label{phiflat}
\end{equation}
Since we are dealing with relativistic neutrinos,
($m_k \ll E_k$), we can approximate,
to the first order, $E_k \simeq E_0 + {\mathrm O}
(m_k^2/(2 \, E_0))$,
where $E_0$ is the energy of a massless neutrino. This
leads to the standard result for the phase shift
\begin{equation}
\Delta \Phi_{kj}^{\mathrm L} \simeq 
\frac{\Delta m_{kj}^2}{2E_0} |\vec x_B - \vec x_A|
= \frac{2\pi L_p(A,B)}{{\mathrm L}_{kj}^{\mathrm osc}} \;,
\label{phaseshift}
\end{equation}
where $\Delta m_{kj}^2$ $= m_k^2 - m_j^2$,
${\mathrm L}_{kj}^{\mathrm osc} = 
(4\pi E_0/\Delta m_{kj}^2)$
is the oscillation length and
$L_p(A,B) = |\vec x_B - \vec x_A|$
is the proper length in the flat space--time.

\end{list}

A generalization to a curved space--time is obtained
by observing that the expression for the phase $\Phi_k$ 
in Eq.(\ref{eq:phase}) is a covariant quantity \cite{sto}:
\begin{equation}
\Phi_k  =
\int_A^P g_{\mu\nu} p_{(k)}^\mu {\mathrm d} x^\nu =
\int_A^P p^{(k)}_\mu {\mathrm d} x^\mu\; ,
\label{eq:covphi1}
\end{equation}
where $p_{(k)}^\mu = m_k {\mathrm d} x^\mu /{\mathrm d} s$
is the four--momentum of $|\nu_k\rangle$,
$g_{\mu\nu}$ is the metric tensor, ${\mathrm d} s$
is the line element and 
$p^{(k)}_\mu = g_{\mu\nu} p_{(k)}^\mu$ is the
conjugate momentum to $x^\mu$. The last expression
in Eq. (\ref{eq:covphi1})
is useful when the metric tensor does not depend on some of the
coordinates $x^\mu$: in that case, the corresponding components 
of $p^{(k)}_\mu$ are constant of motion
along the classical trajectory of the particle, while the 
components of $p_{(k)}^\mu$ are not conserved quantities. 
The momentum $p_{(k)}^\mu$ obeys the mass--shell
condition in the curved metric
\begin{equation}
m_k^2 = g_{\mu\nu} p_{(k)}^\mu p_{(k)}^\nu = 
g^{\mu\nu} p^{(k)}_\mu p^{(k)}_\nu \; .
\end{equation}

Eq.(\ref{eq:covphi1}) denotes the quantum--mechanical phase acquired by a 
particle traveling from $A$ to $B$ in a curved space--time. When 
calculated along the classical path, it corresponds
to the classical action for a free particle.
  
The fact that the time difference between the production 
and detection points is not measured leads to the 
generalization to the curved space--time case of
the light--ray approximation: the phase of
each mass eigenstate must be calculated along the light--ray 
trajectory which links $A$ to $B$
\begin{equation}
\Phi_k^{\mathrm L} = 
\left[
\int_A^B 
\;p^{(k)}_\mu {\mathrm d} x^\mu
\right]_{\mathrm light}\; .
\label{eq:covphiL}
\end{equation} 
This formulation is valid for highly relativistic neutrinos,
since the light--ray approximation is viable only in this case.

\section{Neutrino propagation in the Schwarzschild metric}
\label{sec:curved}

Let us consider the propagation of neutrinos in
the Schwarzschild metric. The line element in the 
coordinate frame $x^\mu=(t,r,\vartheta,\varphi)$ is
\begin{equation}
{\mathrm d} s^2 
=  B(r) {\mathrm d} t^2 -
\frac{{\mathrm d} r^2}{B(r)} -
r^2 {\mathrm d} \vartheta^2 - 
r^2 \sin^2\vartheta {\mathrm d} \varphi^2 \;,
\label{eq:s_metric}
\end{equation}
where $ B(r) = [ 1-(2GM)/r]$,
$G$ is the Newtonian constant 
and $M$ is the mass of the source
of the gravitational field. 
Since the gravitational field is isotropic, the classical orbits 
are confined to a plane, which we choose to be 
the equatorial plane $\vartheta= \pi/2$. Therefore, we have $d\vartheta=0$.
The relevant components of the canonical momentum are 
$p^{(k)}_t$, $p^{(k)}_r$ and $p^{(k)}_\varphi$ 
and they are related by the mass--shell relation as
\begin{equation}
m^2_k =
\frac{1}{B(r)} (p_t^{(k)})^2 - B(r) (p_r^{(k)})^2 -
\frac{(p_\varphi^{(k)})^2}{r^2}\; .
\label{shell}
\end{equation}
Since the metric tensor 
does not depend on the coordinates 
$t$ and $\varphi$, the canonical momenta $p_t^{(k)} \equiv E_k$ and
$p_\varphi^{(k)} \equiv - J_k$ are constant of motion.
With these definitions, the expression of the phase is 
\begin{equation}
\Phi_k^{\mathrm L} =
\hspace{-3pt}
\int_{r_A}^{r_B}
\hspace{-3pt}
\left[
E_k 
        \left(
\frac{{\mathrm d} t}{{\mathrm d} r}
        \right)_{\hspace{-3pt} 0}
\hspace{-3pt} - p_k(r)
-J_k
        \left(
\frac{{\mathrm d}\varphi}{{\mathrm d} r}
        \right)_{\hspace{-3pt} 0}
\right]
\hspace{-3pt}
{\mathrm d} r ,
\label{bigphase}
\end{equation}
where $p_k(r) \equiv - p_r^{(k)}$ and the differentials
$({\mathrm d} t/{\mathrm d} r)_0$ and
$({\mathrm d} \varphi/{\mathrm d} r)_0$ are calculated along
the light--ray path from $A$ to $B$.
It is important to notice that $E_k$ and $J_k$, which are
constants of motion for the geodesic trajectory of $|\nu_k\rangle$,
are not constant along the light--ray trajectory. Instead,
the quantities $E_0$ and $J_0$ for
a massless particle are constant along the light--ray path. 

We briefly comment on
the definition of the energies involved in the 
calculation of $\Phi_k^{\mathrm L}$. 
The constant of motion $E_k$ is conveniently
used in the calculation of the phase 
$\Phi_k^{\mathrm L}$, but it does not directly represent
the energy of the neutrino in the gravitational
field (except for $r=\infty$), since $p^t_k (r) = E_k/B(r)$.
Moreover, the correct definition of the production and detection
energies depends on the reference frames where the physical
processes occur.
For definiteness, 
in our discussion we will choose the local reference frame,
where $E_k^{(loc)} (r) = |g_{tt}|^{-1/2} E_k = B(r)^{-1/2} E_k$.

\subsection{Radial propagation}

For neutrinos propagating in the radial direction, 
${\mathrm d}\varphi = 0$ and 
$({\mathrm d} t/{\mathrm d} r)_0 = \pm B(r)^{-1}$
where the $+$ $(-)$ sign refers to
outward (inward) propagation.
Deriving $p_k(r)$ from the mass--shell relation
of Eq.(\ref{shell}) with $J_k = 0$,
the quantum mechanical phase 
of Eq.(\ref{bigphase}) becomes
\begin{equation}
\Phi_{k}^{\mathrm L}
= \pm
\int_{r_A}^{r_B}
\left(
E_{k}
-
\sqrt{
E_k^2
-
B(r)
m_{k}^2
}
\right)
\frac{{\mathrm d} r }{ B(r) }
\;.
\label{05}
\end{equation}
We can now apply the relativistic expansion using the energy
at infinity $E_k$ as a reference value, i.e. $ m_k \ll E_k $, 
expanding $E_k \simeq E_0 + {\mathrm{O}} ( m_k^2/(2 \, E_0))$,
where $E_0$ is the energy at infinity for a massless particle.
Then, the phase of Eq.(\ref{05}) is easily calculated and
the phase shift which determines the oscillation is
\begin{equation}
\Delta\Phi_{kj}^{\mathrm L}
\simeq 
\frac{\Delta m_{kj}^2 }{ 2 \, E_0 } 
|r_B - r_A|
\;.
\label{deltaphi}
\end{equation}
This result does not 
depend on any assumption on the strength of the
gravitational field.

The expression of the phase shift
in Eq.(\ref{deltaphi}) appears identical
to that of the flat space--time case, Eq.(\ref{phaseshift}).
However, the gravitational effects are implicitly present 
in Eq.(\ref{deltaphi}). When expressed in terms of the
local energy and the proper distance,
Eq.({\ref{deltaphi}) becomes (for a weak field, i.e.
$GM \ll r$)
\begin{eqnarray}
\Delta\Phi_{kj}^{\mathrm L} 
\hspace{-3pt}
&\simeq& 
\hspace{-3pt}
\left(
\frac{ \Delta m^2_{kj} L_p(A,B)}{2 E_0^{(loc)}(r_B)}
\right) \times
\label{phi_measured}
\\
& &
\left[
1
-
G M
\left(
\frac{1}{L_p(A,B)}
\,
\ln \frac{r_B}{r_A}
-
\frac{1}{r_B}
\right)
\right]
\;,
\nonumber
\end{eqnarray}
where $L_p(A,B) \simeq r_B - r_A + GM \ln(r_A/r_B)$
is the proper distance (in the weak field limit).

The first parenthesis on the right--hand side in Eq.(\ref{phi_measured})
is analogous to the flat space--time oscillation phase. The second
parenthesis represents the correction due to the gravitational effects.

The proper oscillation length $L_{kj}^{\mathrm{osc}}$ which is
obtained from Eq. (\ref{phi_measured}) is 
increased because of the gravitational field, as expected.

\subsection{Non--radial propagation}
   
When the classical trajectory is not in the radial
direction, the motion has an additional angular dependence. 
The light--ray differentials are
$({\mathrm d} t/{\mathrm d} r)_0 = 
E_0 B(r)^{-2} p_0(r)^{-1}$
and
$({\mathrm d} \varphi/{\mathrm d} r)_0 = 
J_0 r^{-2} B(r)^{-1} p_0(r)^{-1}$.
The angular momentum $J_k$
is related to $E_k$, the impact parameter $b$ and
the velocity at infinity $v_k^{(\infty)}$ as
$J_k = E_k\,b\,v_k^{(\infty)}$.
Making use of the mass--shell condition 
Eq.(\ref{shell}) and the relativistic expansion $m_k \ll E_k$,
the phase becomes
\begin{equation}
\Phi_k^{\mathrm L}\simeq \pm\frac{m_k^2}{2E_0}
\int_{r_A}^{r_B}
\frac{dr}{\sqrt{
                1-B(r) (b^2/r^2)
               } }\;.
\label{phase-nonr}
\end{equation}
This expression is valid 
for any spherically symmetric (and time--independent) field. 

As a specific example, we consider the propagation of
the neutrino around the massive object. In the weak field
limit, and for the typical situation where $b \ll r_{A,B}$, 
the phase shift is
\begin{equation}
\Delta \Phi_k^{\mathrm L} = 
\frac{\Delta m_k^2}{2E_0}
L_{AB}
\hspace{-3pt}
\left[
1 - \frac{b^2}{2 r_A r_B}
+ \frac{2GM}{L_{AB}}
\right] ,
\label{eq:FinalPhi}
\end{equation}
where $L_{AB} = (r_A + r_B)$.
As discussed for the radial case, a comparison of 
Eq.(\ref{eq:FinalPhi}) with the flat space--time case, 
requires a correct identification of the proper
distance and of the energies involved in the
the physical problem. An interesting application of
Eq.(\ref{eq:FinalPhi}) to the possibility of
gravitational lensing of neutrinos, has been discussed in
Ref.\cite{ourgrav}.

\end{document}